\documentstyle[epsf]{mn}

\newcommand{\oiii}{[O~{\sc iii}]~5007~\AA}

\newcommand{\ciii}{C~{\sc iii}]~1907 \& 1909~\AA}
\newcommand{\degree}{$^{o}$~}
\newcommand{\thetaC}{$\theta^{1}$Ori~C}
\newcommand{\tet}{$\theta^{1}$Ori}
\def\vhel{\ifmmode{V_{{\rm hel}}}\else{$V_{{\rm hel}}$}\fi}
\def\vsys{\ifmmode{V_{\rm sys}}\else{$V_{\rm sys}$}\fi}
\def\kms{\ifmmode{~{\rm km\,s}^{-1}}\else{~km~s$^{-1}$}\fi}

\title[Proplyd jets]
{Two jets from the Orion (M42) `proplyds' -- kinematics, morphologies 
and origins.}
\author[Meaburn, Graham \& Redman]
{J. Meaburn$^{1}$, M. F. Graham$^{1}$ \&
M. P. Redman$^{2}$.\\
$^{1}$Jodrell Bank Observatory, Dept of Physics \& Astronomy, University of
Manchester, Macclesfield, Cheshire SK11 9DL, UK.\\
$^{2}$Department of Physics and Astronomy, University College London,
Gower Street, London WC1E 6BE, UK
}
\date{Received **insert**; in original form **insert**}
\pagerange{\pageref{firstpage}--\pageref{lastpage}}

\begin{document}

\maketitle

\begin{abstract} 
A spatially unresolved velocity
feature, 
with an approaching
radial velocity of $\approx$~100 \kms\ with respect to the systemic
radial velocity,
in a position--velocity array of \oiii\ line profiles
is identified as the kinematical counterpart of a jet
from the proplyd LV~5 (158--323) in the core of the
Orion Nebula. The only candidate in HST imagery for this jet appears 
to be a displaced, ionized knot. 
Also an elongated jet 
projects from the proplyd GMR~15 (161--307). Its receding
radial velocity difference appears at $\approx$~80~\kms\ in the same 
position--velocity array.

A `standard' model for jets from young, low mass stars invokes
an accelerating, continuous flow outwards with an opening angle
of a few degrees. Here an alternative explanation is suggested
which may apply to some, if not all, of the proplyd jets.
In this, a `bullet' of dense material is ejected which ploughs through
dense circumstellar ambient gas. The decelerating tail of material
ablated from the bullet's surface would be indistinguishable
from a continuously emitted jet in current observations.

\end{abstract}


\section{Introduction}

The nature of the compact gaseous knots, dubbed `proplyds'
by O'Dell, Wen \& Hu (1993), in the close vicinity
of the O6 star \thetaC\ in the Orion Nebula, is becoming increasingly
clear. 
They were first discovered (LV 1-6) by Laques \& Vidal (1979) 
in  the optical emission lines with many more identified later as sub-arcsec
diameter 
thermal 
radio sources (Churchwell et al. 1987; Garay 1987;
Garay, Moran \& Reid 1987 - hereafter GMR; Felli et al. 1993a; 
Felli et al. 1993b).
Each proplyd was shown to contain a low mass star (Meaburn 1988;
McCaughrean \& Stauffer 1994) whose youth is suggested by
this partial cocoon of primaeval material (see the dramatic HST
images in O'Dell, Wen \& Hu 1993; O'Dell \& Wong 1996; Bally et
al. 1998). The $\approx$ 50 \kms\ photoevaporated flows, driven by 
the intense Lyman flux 
of \thetaC,
from the ionized 
proplyd surfaces (Meaburn 1988) and their interactions with the
particle wind from \thetaC\ to form stand-off bow-shocks (Hayward, Houck
\& Miles 1994) have most recently been considered by Henney et al. (2002)
and Graham et al. (2002) and references therein.

Jets from young stellar objects (YSOs) were considered as 
one plausible 
explanation of the $\geq$ 100 \kms, highly collimated ($\leq$ 1 \arcsec\ wide)
velocity `spikes' on longslit position-velocity (pv) arrays of \oiii\ line
profiles in ground-based observations of  proplyds 
(Meaburn et al. 1993; Massey \&
Meaburn 1995). The ubiquity of such jets from the proplyds
then became immediately apparent in the HST imagery of 
Bally, O'Dell \& McCaughrean (2000). The HST spectral observations,  
with STIS at 0.1\arcsec\
resolution of the 
\ciii\ profiles from the jet of LV~2,
showed the jet outflow to have a radial velocity extent of $\approx$~120
\kms\ (Henney et al. 2002). This is consistent
with the extent of the velocity spike in the ground-based 1\arcsec\
resolution observations of the \oiii\ 
profiles from LV 2 (fig. 5 in Meaburn et al. 1993; Henney et al. 2002).  
Consequently, it is safe to assume that the narrow velocity spikes
in all of the pv arrays of  \oiii\ profiles observed from the ground in
other proplyds are a measure of their jet velocities.

In the present paper this assumption has been applied to 
two proplyds, LV~5 (158--323) and possibly
GMR~15 (161--307), where
high-speed, collimated, velocity features (`spikes') have been found in 
the pv arrays of \oiii\ profiles
observed from the ground
and where convincing HST images of their jets exist. The bracketed
identifications are from O'Dell \& Wen (1994). 
 
\section{Observations and Results}

The proplyds LV~5 and GMR~15 are sketched against
various features of the Trapezium cluster in the core of
the Orion nebula in Fig.~1.  
\begin{figure*}
\epsfclipon
\centering
\mbox{\epsfbox[0 0 329 327]{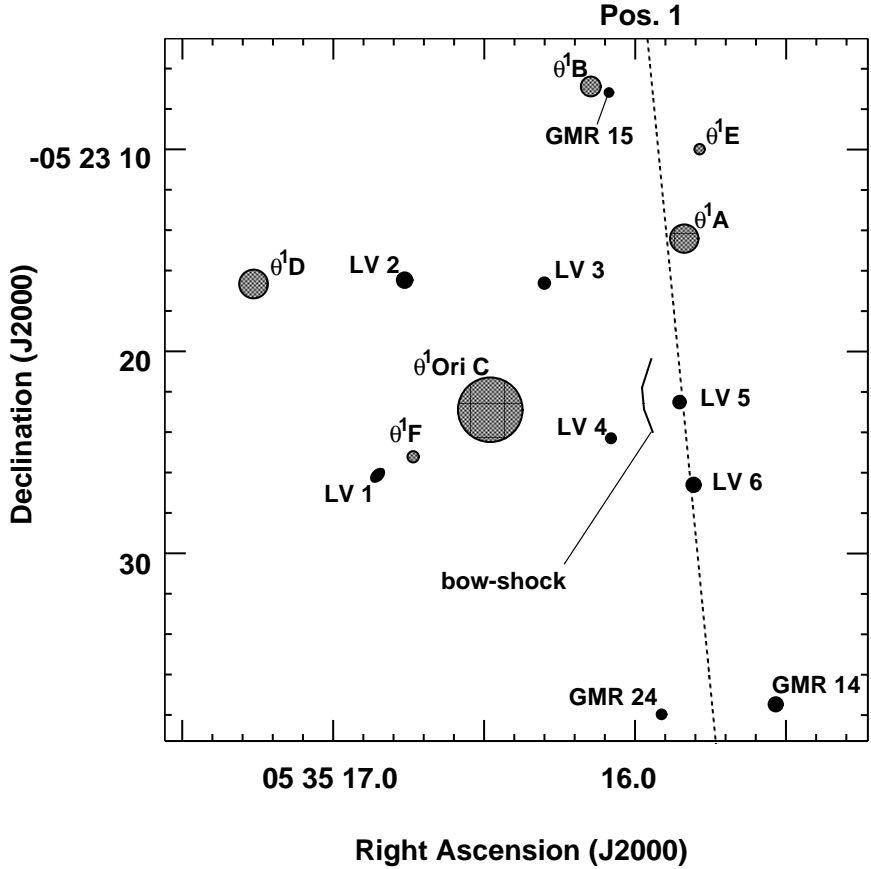}}
\caption{A finding chart of the Trapezium region
of the Orion Nebula, containing \tet\ 
A-F, the six LV knots and objects 14, 15 and 24 from GMR. The slit
position, Pos. 1 is also shown.}
\label{chart}
\end{figure*}
The \oiii\ line profiles were obtained with the Manchester
echelle spectrometer (MES - Meaburn et al. 1984) 
combined with the Isaac Newton Telescope (INT) at various
times between 1991 and 1993. Full details of the 
observational setup are reported in Massey \& Meaburn (1995).
The most notable aspect of the observation along slit
length marked Pos. 1 in Fig.~1 was that
sub-arcsec `seeing' prevailed. A resolution
of 0.9\arcsec\ was achieved in practice.

The pv array of \oiii\ line profiles is compared in Fig.~2
with a section of the HST \oiii\ archive image (from the GO 5469
programme of J. Bally). The MES slit orientation and width is
indicated. 
\begin{figure*}
\epsfclipon
\centering
\mbox{\epsfxsize=\textwidth\epsfbox[0 0 722 366]{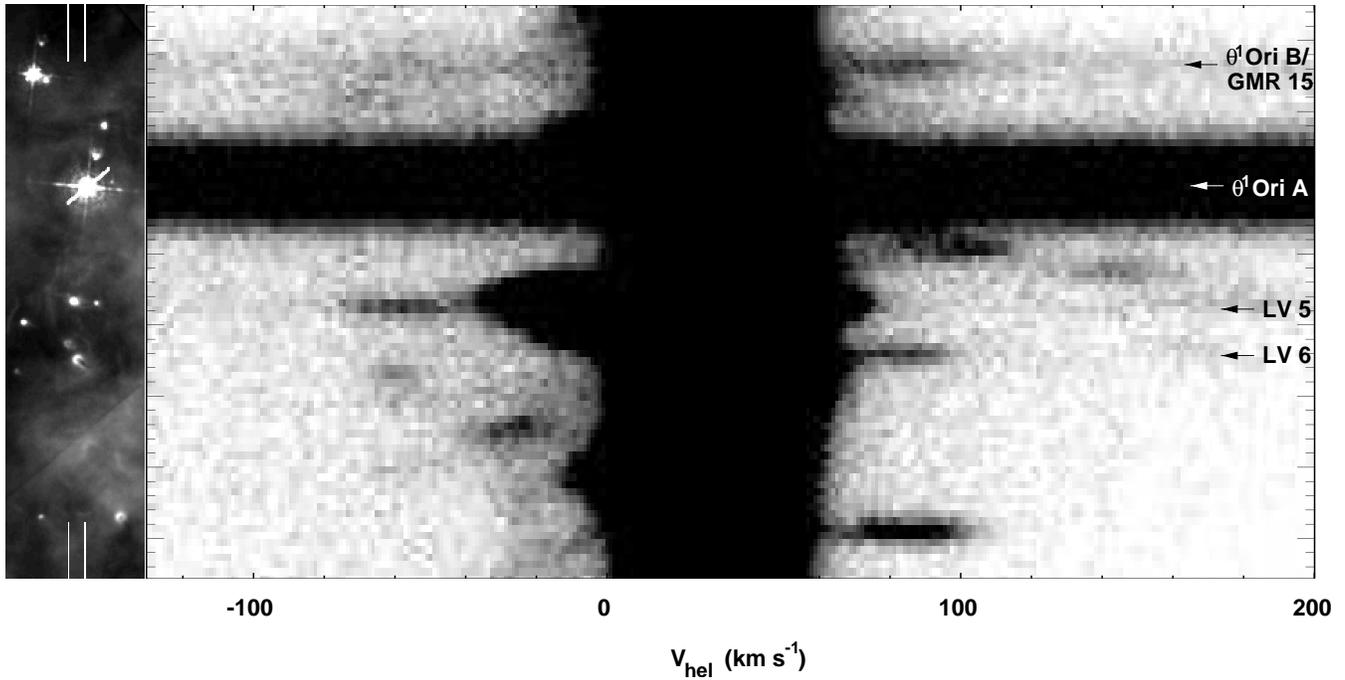}}
\caption{The \emph{HST} \oiii\ image (left hand panel) 
of the region surrounding
Pos. 1 in Fig.~1 aligned with the \oiii\ pv array of this
slit position. The image is centered on Pos. 1 and the  white lines
indicate a width of 1\arcsec, which corresponds to the slit width of
0.4\arcsec\ convolved with seeing of 0.9\arcsec. The vertical extent
of the pv array (along the slit) is 41\arcsec.} 
\label{hst_pv}
\end{figure*}
For LV~5 an attempt has been
made to isolate the \oiii\ emission line profile from the proplyd from
the strong emission line profile of the nebular background. This was
achieved by extracting the \oiii\ profile which included
the proplyd from an 0.9\arcsec\ length of the pv array. The 
mean of two profiles from adjacent and equal lengths
of the slit, above and below this position, after 
interpolation, was then subtracted.
The results of this procedure for LV 5 are shown
in Fig.~3. 
\begin{figure*}
\epsfclipon
\centering
\mbox{\epsfxsize=\textwidth\epsfbox[0 0 758 501]{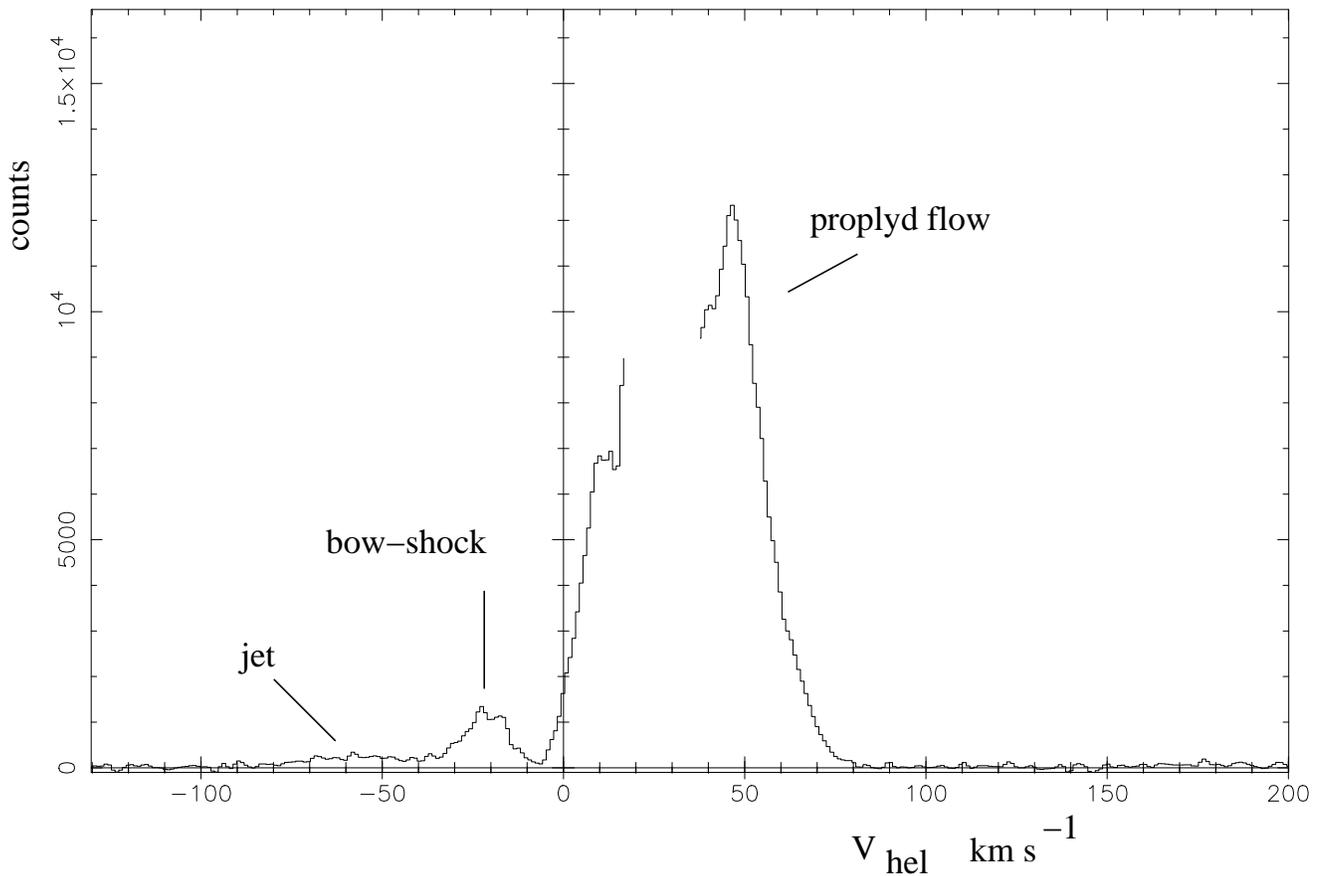}}
\caption{An \oiii\  profile taken with the MES of the proplyd
LV~5.  The nebula background
has been subtracted by interpolating adjacent regions either side of
LV 5 along the slit length.}
\label{profile}
\end{figure*}
The only uncertain parts
of this profile are from the conservatively
large heliocentric radial velocity range
\vhel\ = 15 to 35 \kms\ which contains the intense \oiii\ emission
around the systemic heliocentric  radial velocity 
(\vsys\ $\approx$\ 25~\kms) of M42. 
  
The HST \oiii\ archive images of LV 5 and GMR 15, along with its jet,
j160--307 (Bally et al. 2000) are shown in Figs.~4 and
5 respectively for comparison.
\begin{figure*}
\epsfclipon
\centering
\mbox{\epsfbox[0 0 363 531]{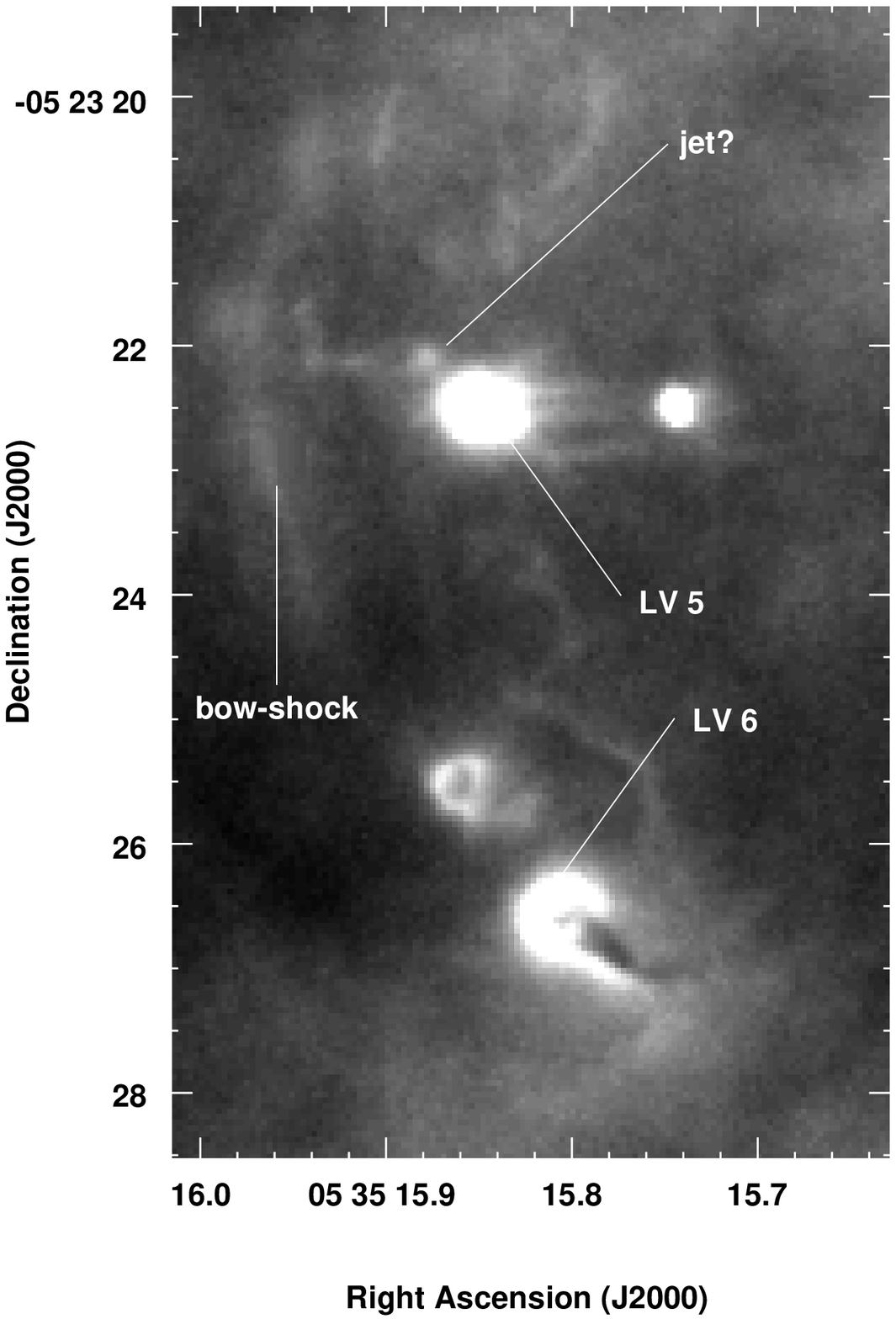}}
\caption{\emph{HST} \oiii\ image of LV 5 and LV 6.  LV 5's possible
jet is indicated along with the bow-shock due to the interaction
of the proplyd flow and the wind of \thetaC.}
\label{lv5_lv6}
\end{figure*}
\begin{figure*}
\epsfclipon
\centering
\mbox{\epsfbox[0 0 372 355]{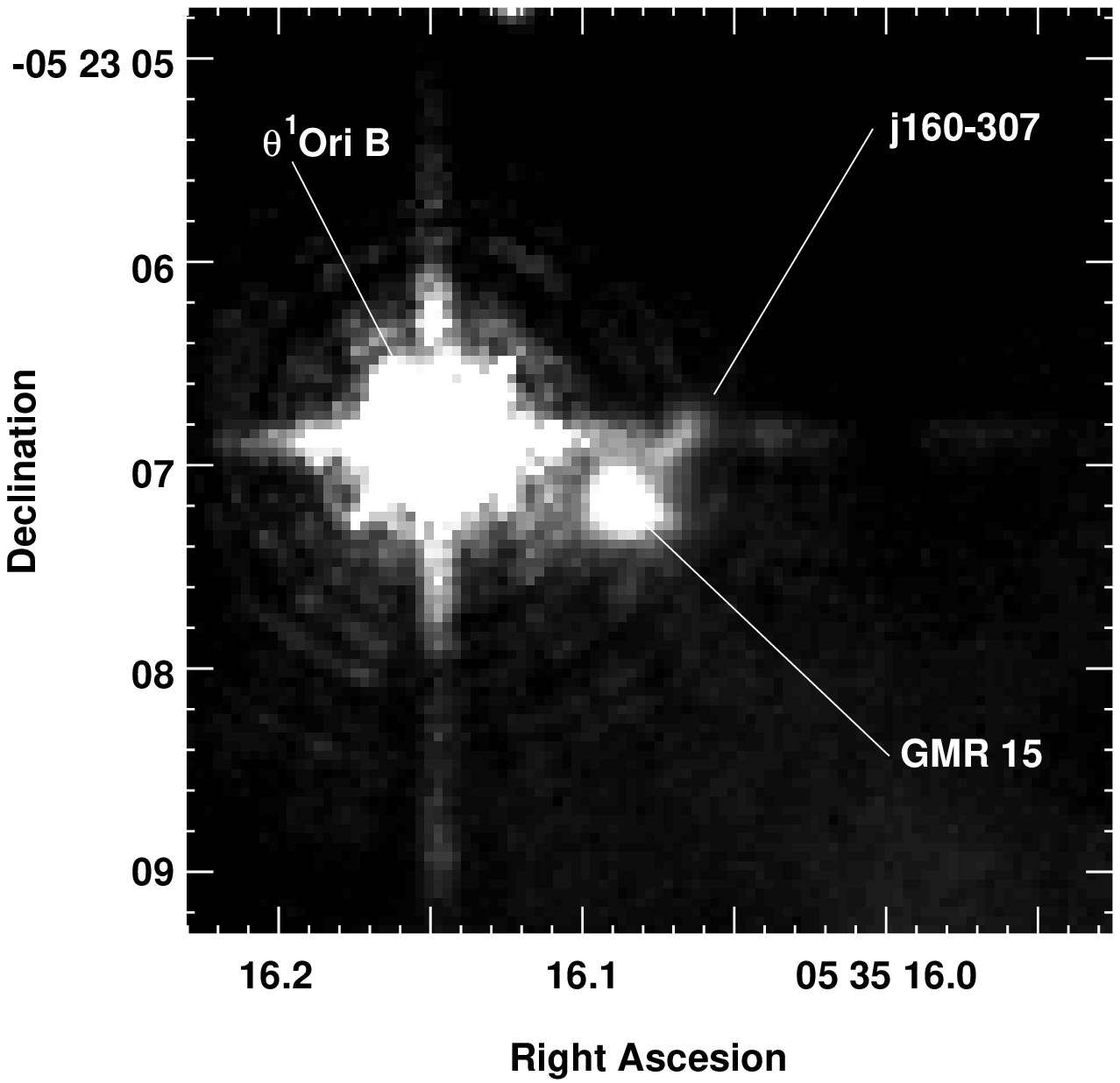}}
\caption{\emph{HST} \oiii\ image of the star \tet\ B and  
proplyd GMR 15 along with its
jet, j160--307.}
\label{j160_307}
\end{figure*}
Here the jet candidates are identified.
The velocity component marked `bow-shock' in Fig.~3 is
coincident with 
the similarly spatially extended feature in the pv array in Fig.~2
that must originate in the arc of emission (see Fig.~4)
towards \thetaC. It is reasonable to assume that this arc
delineates the bow-shock as the particle wind of \thetaC\ 
encounters the obstruction of LV~5 and its photo-evaporating gas. 

The high-speed `spike' from \vhel\ = -40 to -80 \kms\ in the pv array in 
Fig.~2 (and see its identification
in the profile in Fig.~3) is convincingly that expected of
a jet from LV~5  
whose width is unresolved by the 0.9\arcsec\ resolution of these ground-based
observations. The knot adjacent to LV~5,
identified in the HST image in Fig.~4,
is the only available candidate for such a  jet. Several of the
jets identified by Bally et al. (2000) had the
appearance of similar faint emission line knots.

The positive velocity `spike' in Fig.~2, closely associated
with the proplyd LV~6, is similarly narrow and extends to
\vhel~$\approx$ 100 \kms. However, there is no corresponding
jet-like feature in the HST image. The broader `spike' in
Fig.~2 where
the slit passes between knot 14 and 24 is less convincingly of
proplyd origin even though it extends  to \vhel\ $\approx$ 120 \kms.
Here the slit intercects a filament which may have a Herbig-Haro like
nature (Canto et al. 1980).

However, a trace of the velocity `spike' of the very positively 
identified jet, j160--307, shown in Fig.~5, could be
the feature in Fig.~2 which extends 
out to \vhel\ $\approx$ 100 \kms. The slit in Fig.~2 must just 
graze this proplyd for the continuous spectrum of the adjacent
star $\theta^{1}$Ori~B can be seen in the pv array.

\section{Discussion}

In an attempt to understand the nature of the proplyd jets, one can
compare them to the molecular jets seen in low mass star forming
regions. Some of these molecular outflows have also been observed to
be characterised by a `Hubble-like' velocity law (see e.\@g.\@ Lada \&
Fich 1996) such that the highest velocity gas is found furthest from
the star implying an acceleration outwards along the length
of the jet. This has the implication that all of the gas has the same
dynamical timescale. Acceleration mechanisms for such outflows are
unclear at present but the most popular possibility is that the
molecules are accelerated in jet-driven bow-shocks (Downes \& Ray
1999) though there are problems with most of the current models (Lada
\& Fich 1996). A straight-forward view would then be that the proplyd
jets are simply continuous, nearly collimated, outflows from the YSO
to the jet tips with jet opening angles of $\leq$~10\degree\ if they
are unresolved by the HST 0.1\arcsec\ resolution and if they have a
length of 0.5\arcsec. See Meaburn \& Dyson (1987) for such a model
applied to HH 46-47 and its
first application to proplyd jets by Bally et al (2000). 
Such models would of course have to be modified,
mainly to account for the highly ionizing environment, but also
because the mass and momentum flux of these jets are rather feeble
compared with that from HH 46-47 (Henney et al 2002). The jet material
will be ionized unless it is
shielded from $\theta ^1$Ori~C by the proplyd disk and Raga et al
(2000) have shown that in fact all the jet material will be promptly
ionized in this way so that a neutral core within the jet is
unlikely. The image of such a jet from an Orion proplyd would be
indistiguishable from that from GMR 15 in Fig.~5 because its declining
width towards the proplyd YSO would be unresolved by HST.

As ballistic high-speed `bullets' are associated 
with several circumstellar eruptive phenomena [see Lopez, Meaburn \&
Palmer (1997), Bryce et al (1997) and O'Connor et al (2000) for 
PNe and Redman, 
Meaburn \& Holloway (2002)
for the Luminous Blue Variable, Eta Carinae] an alternative
explanation of the proplyd jets is worth considering. 
They could be a consequence of bullets with ablated flows. Incidentally,
although such bullet phenomena are observationally well
established in a variety of circumstellar
environments their origins are far from being understood.  

\subsection{The bullet/ablated trail model}

The jet candidate appears as a single knot adjacent to LV~5 in
Fig.~4 (and in other examples in Massey \& Meaburn 1995 and Bally et
al. 2000) an explanation as a collimated outflow 
is not too convincing (unless only the bow-shocked gas
at the tip of a collimated monopolar jet is being seen). However, a
simple variant on the model involving a collimated, continuous
outflow is suggested that may be
applicable in some proplyd jet systems. Instead of the jet being
generated by continuous mass loss from the central system, a single
brief mass ejection event from the YSO is postulated. This will give
rise to a dense clump of ejecta material which will behave like a ballistic
bullet since it will not be continually accelerated. In this model for
a proplyd jet it is assumed that photo-ionized material is ablated
(Dyson, Hartquist \& Biro 1993; Klein, McKee \& Colella 1994) 
off the bullet head as it ploughs
through ambient gas leaving a trail in its wake. The trail will
broaden at approximately the sound speed and be decelerated by oblique
shocks. This could appear in the HST images as a compact knot of
photoionised emission possibly with an elongated collimated feature
{\it decelerating} towards the YSO but again resembling, at
0.1\arcsec\ resolution, the jet protruding from GMR~15 in Fig.~5.
Within this explanation the distinction between the GMR~15 
elongated jet 
and LV~5 compact knot could simply be that the bullet causing
the extended GMR~15  feature is ploughing through denser
circumstellar gas causing more ablation.
The
decelerating `strings' of high-speed gas, preceded by bullets from the
explosive LBV star, $\eta$ Carinae seem to be particularly
well described by such a
model (Redman, Meaburn \& Holloway, 2002) and motivates this suggestion
although the ejection speeds involved in the proplyd phenomenon are 
$\approx$~10 times lower. 

The apparent length of the proplyd jets are $\sim 0.5$ arcsec, giving them
a physical length of $\sim 200~{\rm au}$ or $3\times 10^{16}~{\rm
cm}$. If the bullet is travelling at $\sim 120~{\rm km~s^{-1}}$ and
has not slowed down significantly then it was ejected from the proplyd
approximately 10 years ago. Repeated HST observations should clearly
detect the motion of these features on a timescale of just a few years.
The jet width is not resolved by the HST at distance of the Orion nebula,
450pc, which implies that the diameter of the jet is $< 45~{\rm au}$
or $7\times 10^{14}~{\rm cm}$. If the jet represents a decelerating
wake behind a bullet then it is unlikely that the bullet is
significantly wider than the wake. 

Within this bullet/ablated flow model,the maximum length of the 
trail is controlled by how far the bullet is
able to travel into the ambient medium before being destroyed. The
destruction length depends on the density contrast between the clump
and ambient medium $\xi=n_{\rm c}/n_{\rm i}$. The emission measure of
jet j160-307 (measured by Bally et al 2000) leads to a density
estimate of $n_{\rm e}=1.3\times 10^5~{\rm cm^{-3}}$ for the jet
material. This corresponds to a density contrast of $\xi \sim 50$
between this material and the background density in Orion. Thus the
original bullet should have been denser by a factor of at least this
much. Henney et al (2002) using the models of Raga et al (2000) point
out that any condensation dense enough to remain neutral while being
exposed to the ionizing radiation from $\theta ^1$Ori~C will have a
surface brightness comparable to the proplyd surface since in both
cases the emission will be from gas that has just passed through a
D-type ionization front. This does not rule our model out because the
density required for the ionization front to be D-type is $2\times
10^{8}~{\rm cm^{-3}}$ or $\xi\sim 10^3$. This is far greater than that
expected from the bullet in our model and thus the bullet is likely to
be ionized throughout. Furthermore, the ionization front will have
been the fast R-type with a propagation velocity of $\sim 10^{4}~{\rm
km~s^{-1}}$ so that the bullet would have been ionized very soon after
being ejected from the proplyd.

There are two clear characteristics that should distinguish a bullet
and ablated
trail from its continous outflow competitor. Firstly the former
will broaden towards the YSO and the latter become narrow. Secondly,
the ablated trail would exhibit deceleration along its length towards
the YSO and, if spatially  resolved in this direction, 
show as a linear change in radial velocity
towards the YSO. This characteristic would appear, if the trail/jet is
unresolved as in the present ground based spectral 
observations in Fig.~2, as a
broad velocity spike down to \vsys\ in a pv array of line
profiles. The continuous non-accelerating jet, in contrast, 
would have a reasonably
uniform speed along its length and appear as a narrow, isolated,
high-speed feature, displaced from \vsys, in such pv arrays. STIS
observations down the length of the jet could discriminate between
these possibilities.

\section{Conclusions}

An elongated jet in an HST image 
projecting from proplyd GMR 15
appears as a velocity `spike' in the corresponding
pv array obtained with MES. 
Also, the only jet candidate to explain the high-speed unresolved
feature in an MES pv array is found to be a displaced ionized knot in
the HST imagery of the proplyd LV~5.
 
Two distinctly different dynamical explanations are considered
for these jets.
In many cases standard jet models (suitably modified for the
conditions within the Orion Nebula) may be able to describe proplyd
jets. However, we suggest that in some instances it may be more useful
to model them as being due to the passage of discrete bullets of
ejecta that are promptly ionized and that are ablated as they travel
through the ambient medium. The trailing ablated material 
is then identified as
the jet, with the bullet at its head. Such a dynamical mechanism has
the attraction that the `Hubble-type' velocity law, if
found in future observations of the proplyd jets, is naturally
explained since material that was ablated earlier has slowed down more
than material more recently incorporated into the flow. Kinematic
differences between this model and a continous jet model could be
investigated using STIS spectroscopy with its 0.1\arcsec resolution.

\section*{Acknowledgements} MPR and MFG are supported by a PPARC
Research Assistantship and a Research Studentship respectively.  

\bibliographystyle{mnras}

\begin{thebibliography}{}

\bibitem[]{}
Bally~J., Sutherland~R.~S., Devine~D., \& Johnstone~D. 1998, AJ, 116, 
293.\\

\bibitem[]{}
Bally J., O'Dell C. R., \& McCaughrean M. J. 2000, AJ, 119, 2919.\\

\bibitem[]{}
Bryce, M., L{\'o}pez, J. A., Holloway, A. J. \& Meaburn, J.
1997, ApJ, 487, L161.\\

\bibitem[]{} 
Cant{\'o} J., Goudis C., Johnson P.~G., \& Meaburn J.\ 1980, A\&A, 
85, 128.\\

\bibitem[]{}
Churchwell E., Felli M., Wood D. O. S., \& Massi M. 1987, ApJ, 321, 516.\\

\bibitem[]{}Downes T.P.,  Ray T.P.,  1999, A\&A, 345, 977.\\

\bibitem[]{} Dyson J.~E., Hartquist T.~W., \& Biro S.\ 1993,
MNRAS, 261, 430.\\

\bibitem[]{}
Felli M., Churchwell E., Wilson T. L., \& Taylor G. B. 1993, 
A\&ASS, 98, 137.\\

\bibitem[]{}
Felli~M., Taylor~G.~B., Catarzi~M., Churchwell~E., \& Kurtz~S. 1993,
A\&ASS, 101, 127.\\ 

\bibitem[]{} 
Garay G.\ 1987, Revista Mexicana de Astronomia y Astrofisica,
vol.~14, 14, 489.\\

\bibitem[]{}
Garay~G., Moran~J.~M., \& Reid~M.~J. 1987, ApJ, 314, 535 (GMR)\\

\bibitem[]{}
Graham M. F., Meaburn J., Garrington S. T., O'Brien T. J., Henney
W. J., \& O'Dell C. R. 2002, ApJ, 570, in press.

\bibitem[]{}
Klein R. I., Mckee C. F. \& Colella P. 1994, ApJ, 420, 213.\\

\bibitem[]{} Hayward T.\ L., Houck J.\ R., \& Miles J.\ W. 1994,
ApJ, 433, 157

\bibitem[]{} 
Henney W.~J., O'Dell C.~R., Meaburn J., Garrington S.~T., \&
Lopez J.~A.\ 2002, ApJ, 566, 315

\bibitem[]{}
Lada C.J.,  Fich M.,  1996, ApJ, 459, 638\\

\bibitem[]{}
Laques P., \& Vidal J. P. 1979,  A\&A, 73, 97

\bibitem[]{}
L{\'o}pez J. A., Meaburn, J. \& Palmer, J. W. 1997, ApJ, 415, L135.

\bibitem[]{}
Massey R. M., \& Meaburn J. 1995, MNRAS, 273, 615

\bibitem[]{}McCaughrean M.\ J., \& Stauffer J.\ R.\ 1994, AJ,
108, 1382.\\

\bibitem[]{} 
Meaburn J., Blundell B., Carling R., Gregory D.\ F., Keir D., \& Wynne 
C.\ G. 1984, MNRAS, 210, 463.\\

\bibitem[]{} Meaburn J.~\& Dyson J.~E.\ 1987, MNRAS, 225, 863.\\ 

\bibitem[]{}
Meaburn J. 1988, MNRAS, 233, 791.\\

\bibitem[]{} Meaburn J., Massey R. M., Raga A. C., \& Clayton
  C.A. 1993, MNRAS, 260, 625.\\

\bibitem[]{}
O'Connor J. A., Redman M. P., Holloway A. J., Bryce M.,  L{\'o}pez J. A.
~\& Meaburn J. 2000, ApJ, 531, 336.

\bibitem[]{}
O'Dell C. R., Wen Z., \& Hu X. 1993, ApJ, 410, 696.\\

\bibitem[]{}
O'Dell C. R. \& Wen Z. 1994, ApJ, 436, 194.\\

\bibitem[]{}
O'Dell~C.~R., \& Wong~K. 1996, AJ, 111, 846.\\

\bibitem[]{}
Raga A.C., L{\o}pez-Mart{\'\i}n L., Binette L., L{\'o}pez J.A.,
Cant\'{o} J., Arthur S.J., Mellema G., Steffen W., Ferruit P., 2000,
MNRAS, 314, 681.\\

\bibitem[]{}
Redman M. P., Meaburn J., \& Holloway A. J. 2002, MNRAS, 332, 754.

\end{thebibliography}

\end{document}